\documentclass[twocolumn]{aastex6}
\RequirePackage{lineno}
\usepackage{amsmath}
\usepackage{amssymb}
\usepackage{hyperref}
\usepackage{amsthm}
\usepackage{natbib,aasdefs,url,bm}
\usepackage{array}
\usepackage{float}
\usepackage{graphicx}
\usepackage{subfigure}
\usepackage{color}
\usepackage{lineno}

\newcounter{ichi}
\newcounter{ni}
\newcounter{san}
\newcounter{yon}
\def\be{\begin{equation}}
\def\ee{\end{equation}}
\def\ba{\begin{eqnarray}}
\def\ea{\end{eqnarray}}

\def\yuanccs{}
\def\yuancc{}
\def\yuan{}

\slugcomment{}

\shorttitle{Secondary Radio and X-ray Emissions from Galaxy Mergers}
\shortauthors{Yuan, Murase and M\'esz\'aros}

\begin{document}

\title{Secondary Radio and X-ray Emissions from Galaxy Mergers}
\author{Chengchao Yuan\altaffilmark{1}}\email{cxy52@psu.edu}
\author{Kohta Murase\altaffilmark{1,2} }
\author{Peter M\'esz\'aros\altaffilmark{1}}
\altaffiltext{1}{Department of Physics; Department of Astronomy and Astrophysics; Center for Particle and Gravitational Astrophysics, The Pennsylvania State University, University Park, PA 16802, USA}
\altaffiltext{2}{Center for Gravitational Physics, Yukawa Institute for Theoretical Physics, Kyoto University, Kyoto 606-8502, Japan}


\begin{abstract}
Shocks arising in galaxy mergers could accelerate cosmic-ray (CR) ions to TeV-PeV energies. While propagating in the intergalactic medium, these CRs can produce high-energy neutrinos, electron-positron pairs and gamma-rays. 
In the presence of intergalactic magnetic fields, the secondary pairs will radiate observable emissions through synchrotron radiation and inverse Compton scattering. In this paper, we demonstrate that these emissions can explain the radio and X-ray fluxes of merging galaxies such as NGC 660 and NGC 3256. 
Using our model in combination with the observations, we can constrain the gas mass, shock velocity, magnetic field and {\yuan the CR spectral index $s$} of these systems. 
For NGC 660 a single-zone model {\yuan with a spectral index $2.1\lesssim s\lesssim2.2$} is able to reproduce simultaneously the radio and X-ray observations, while a {\yuan simple one-zone scenario with $s\sim2$ can describe the radio and a large fraction of X-ray observations of NGC 3256.} 
Our work provides a useful approach for studying the dynamics and physical parameters of galaxy mergers, which can play an important part in future multi-messenger studies of similar and related extragalactic sources.
  
\end{abstract}
\keywords{cosmic rays --- galaxies: interactions --- galaxies: individual (NGC 660, NGC 3256) --- radio continuum: galaxies --- X-rays: galaxies }

\section{Introduction}
\label{sec:intro}
Star-forming galaxies including starbursts have been considered as possible reservoirs of cosmic rays (CRs) and sources of associated neutrinos and gamma rays \citep[e.g.,][]{Loeb:2006tw,thompson2007starburst,Murase:2013rfa}, in which the CRs can be supplied by not only supernovae but also hypernovae, superbubbles and active galactic nuclei \citep[][]{senno2015extragalactic,Xiao:2016rvd,Tamborra:2014xia,2016NatPh..12.1116W,2017A&A...607A..18L,Liu:2017bjr}.   
Interacting galaxies, which may be accompanied by starburst activities, have also been considered as additional accelerators of CRs \citep{kashiyama2014galaxy,yuan2018cumulative}. 
Under the conditions typical of galaxy merger systems synchrotron emission can extend from the radio band to the X-ray regime, while the inverse Compton scattering may be important in the ultraviolet (UV) and up to beyond the X-ray band. 

In this work we formulate a model which is capable of reproducing the radio and X-ray observations of specific systems using synchrotron and synchrotron self-Compton (SSC) or external inverse Compton (EIC) emissions from high-energy secondary electron-positron pairs produced by the CR interactions in such systems. Here the EIC is caused by scatterings with the cosmic microwave background (CMB), {\yuan infrared/optical starlight (SL)} and extragalactic background light (EBL). 
In addition, since the radiation spectrum of the merging galaxies is determined by the dynamics of the galaxy interactions and the resulting physical conditions, this enables us to provide constraints on the magnetic field $B$, shock velocity $v_s$, gas mass $M_{\rm g}$, etc. Different from \cite{lisenfeld2010shock} where shock-accelerated electrons are employed to describe the radio emissions of two colliding galaxies, UGC 12914/5 and UGC 813/6, we present an alternative model based on the secondary emission from inelastic $pp$ collisions to reproduce simultaneously the radio and X-ray observations of NGC 660 and NGC 3256. {\yuancc In general, secondary electrons are more natural to explain the electromagnetic emissions in merging galaxies. For the observed CRs, the electron acceleration efficiency, the fraction of plasma energy deposited to electrons, is at least two orders lower than the proton acceleration efficiency, e.g.  $K_{e/p}=\epsilon_e/\epsilon_p\sim{10}^{-4}-10^{-2}$ \citep{Jones2011,morlino2012strong}. This value is also consistent with the observations of Galactic supernova remnants. Furthermore, the recent particle-in-cell simulation shows a similar value, $K_{e/p}\simeq10^{-3}$ \citep{katz2008shell,caprioli2014simulations,park2015simultaneous}. The ratio of the primary electrons (from shock accelerations) and the secondary electrons and positrons is approximately 
\begin{linenomath*}
\[
\frac{\mathcal E_{e\rm, primary}}{\mathcal E_{e,\rm sec}}\simeq\frac{6\epsilon_e}{\min[1,f_{pp,\rm g}]\epsilon_p}\lesssim10^{-1}.
\]\end{linenomath*}
where $f_{pp,\rm g}$ is the effective $pp$ optical depth in the emitting region.
Therefore in our model with the typical electron/proton acceleration efficiencies, emission from primary electrons directly accelerated in shocks is {subdominant} compared to that from secondary electrons and positrons from $pp$ collisions and pion decays. This is consistent with \cite{murase2018high} where they suggest that the secondary emissions overwhelm the primary component in nearly proton calorimetric sources.  {\yuanccs It is possible that primary electrons can provide a non-negligible contribution if $K_{e/p}\gtrsim0.1$, considering that $K_{e/p}$ is poorly constrained theoretically and observationally for this system.} In the following text, we focus on the primary electron/positron scenario and omit the primary electron contribution.}

As a well-studied interacting system, NGC 660 is a galaxy formed by the collision of two galaxies \citep{van1995polar}, which has been observed in both radio \citep[e.g.,][]{douglas1996texas,large1981molonglo,condon2002radio,condon1998nrao,dressel1978arecibo,bennett1986green,becker1991new,gregory199187gb,sramek19755}, microwave, infrared, UV and X-ray \citep[e.g.,][]{fraternali2004further,liu2010chandra,brightman2011xmm,white2000wgacat} bands. Also, the magnetic field in the core region of NGC 660 is constrained in the range of $16\pm5\ \mu {\rm G}$ through polarization studies \citep{drzazga2011magnetic}. In this paper, we take NGC 660 as an example and use our model to reproduce the radio, UV and X-ray fluxes. We also apply our model to constrain the shock velocity and gas mass of the core region of NGC 660 by using the magnetic field $16\pm5\ \mu {\rm G}$ as a precondition. To show that our model's applicability can be extended to other similar systems, we also consider another well-studied galaxy formed through a merger, NGC 3256, as a supplementary template.

This paper is organized as follows. In \S\ref{sec:elec_input}, we formulate the secondary electron-positron spectrum and calculate resulting electromagnetic emissions, including synchrotron radiation and SSC/EIC components. 
In \S\ref{sec:application} , we apply the formalism in \S\ref{sec:elec_input} to the core regions of NGC 660 and NGC 3256. 
A summary and discussion, including comparison with previous work in the context of starburst galaxies, is given in \S\ref{sec:summary}. 

\section{Secondary electron spectrum and electromagnetic emissions}
\label{sec:elec_input}

The pions produced in the $pp$ collisions between shock-accelerated CR ions and the galaxy gas generate, besides high-energy neutrinos and $\gamma$-rays, also copious quantities of high-energy electron-positron pairs. These high-energy leptons may produce observable synchrotron emissions while propagating inside the galactic magnetic fields. Here, considering the conservation of lepton numbers and muon decays, we approximate the total electron-positron injection spectrum to be the same with the neutrino production spectrum. Following the procedure in  \cite{yuan2018cumulative}, the electron injection spectrum can be written as
\begin{linenomath*}
\begin{equation}
\begin{split}
\varepsilon^{2}\mathcal N_e(\varepsilon)=&\frac{1}{3}\varepsilon^{2}\frac{dN_\nu}{d\varepsilon}=\frac{1}{12}\epsilon_p\mathcal{C}^{-1} M_{\rm g}v_{\rm s}^2\\
		 &\times\min\left[1,f_{pp,\rm g}\right]_{\varepsilon_p\simeq20\varepsilon},
	\end{split}
\label{eq:elec_injection}
\end{equation}
\end{linenomath*}

\begin{figure}
\includegraphics[width=0.5\textwidth]{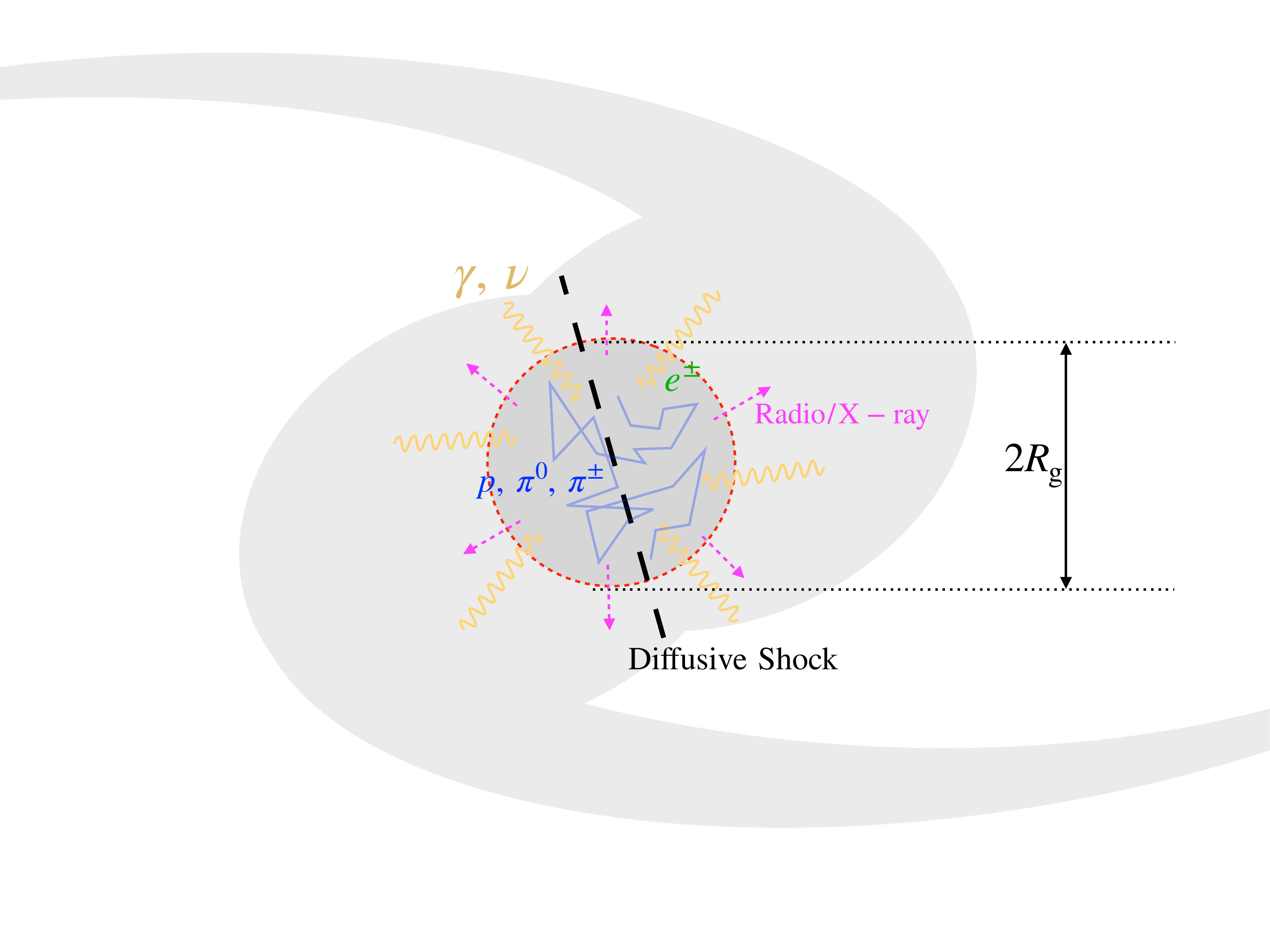}
\caption{Schematic figure showing the merger of two galaxies. The shock was simplified as a straight line across the dense core region. It is also in the core region where interactions occur and neutrinos as well as electromagnetic radiation are produced.}
\label{fig:merger}
\end{figure}

where $\epsilon_p$ is the CR ion acceleration efficiency {\yuan{(normally fixed as 0.1)}}, $\mathcal C=\ln({\varepsilon_{p,\rm max}/\varepsilon_{p,\rm min}})$ is the normalization coefficient for a $\varepsilon^{-2}$ spectrum, $M_{\rm g}$ is the gas mass of the merging region, $v_s$ is the shock/collision velocity and $f_{pp,\rm g}=\kappa_{pp} c n_{\rm g} \sigma(\varepsilon_p)\min[t_{\rm esc},t_{\rm dyn}]$ is $pp$ optical depth inside the galaxy. In this expression, $\kappa_{pp}=0.5$ is the proton inelasticity, $c$ is the speed of light, $n_{\rm g}$ is the gas density, $t_{\rm esc}$ is the escape time of CRs, $t_{\rm dyn}\simeq R_{\rm g}/v_{\rm s}$ is the dynamic time of the merger and $\sigma(\varepsilon_p)$ is the $pp$ cross section given by \cite{kafexhiu2014parametrization}. {\yuancc As galaxies merge, strong shocks occur with a complicated morphology over a galaxy scale, while merging cores of the two galaxies lead to a dense core region. Particles are accelerated by the shocks, and then will be distributed in a galaxy scale. The CRs diffusing in the core region will make neutrinos and gamma rays efficiently.  In this work, as a simplified approximation without covering the details of the shock structure, we assume that shocks are CR accelerators, which inject high energy CRs to the core region of the merging systems and initiate subsequent interactions. Figure \ref{fig:merger} shows the schematic. After leaving the accelerator, the particles can propagate diffusively or get advected away through galactic winds,} therefore the net escape rate is the sum of diffusion rate and advection rate, e.g. $t_{\rm esc}^{-1}\approx t_{\rm diff}^{-1}+t_{\rm ad}^{-1}$. Although the maximum CR energy $\varepsilon_{p,\rm max}$ and effective $pp$ optical depth $f_{pp,\rm g}$ depend on the geometry of the colliding galaxies, for simplicity and consistency, we assume that the neutrinos are produced inside the core region of the interacting system and calculate the electromagnetic radiation therein. This hypothesis is in good agreement with the radio maps of NGC 660 and NGC 3256.  Hence, to fully depict the physical condition of the core region, we introduce several quantities, the radius $R_{\rm g}$, the average magnetic field $B$ as well as the previously defined gas mass $M_{\rm g}$ and shock velocity $v_s$. Using these parameters, we can write down the maximum CR energy, gas density and diffusion time explicitly as $\varepsilon_{p,\rm max}=\frac{3}{20}eB_sR_{\rm g}\frac{v_s}{c}$ \citep{drury1983introduction}, $n_{\rm g}=M_{\rm g}/(\frac{4}{3}\pi m_pR_{\rm g}^3)$ and $t_{\rm diff}=R_{\rm g}^2/(6D_{\rm g})$, respectively.  Here, $m_p$ is the proton mass, $D_{\rm g}$ is the diffusion coefficient and $B_s$ is the post-shock magnetic field which can be parameterized as a fraction of the ram pressure $B_s^2/8\pi=\frac{1}{2}\epsilon_Bn_{\rm g}m_pv_s^2$ \citep{kashiyama2014galaxy}. As for the diffusion coefficient $D_{\rm g}$, we use a combined large and small angle diffusion equation as in \cite{senno2015extragalactic,casse2001transport} and \cite{yuan2018cumulative} and then {\yuan it can be written explicitly as
\begin{linenomath*}
\begin{equation}\begin{split}
t_{\rm diff}\simeq &4.28 {\rm\ Myr}\left(\frac{R_{\rm g}}{3\ \rm kpc}\right)^2\left(\frac{D_0}{10^{29}\ {\rm cm^2\ s^{-1}}}\right)^{-1}\\
&\times\left[(\varepsilon/\varepsilon_{c,\rm g})^{1/2}+(\varepsilon/\varepsilon_{c,\rm g})^{2}\right]^{-1}
\end{split}
\end{equation}
\end{linenomath*}
{\yuanccs where $D_0$ is defined by $D_0\simeq cl_c/20$, $l_c\simeq 0.1R_{\rm g}$ is the coherence length of the magnetic field fluctuations and $\varepsilon_{c,\rm g}\approx eBl_c$ is the characteristic energy.} As for the advection, the typical values of wind velocity in star-forming galaxies and star burst galaxies range from 500 $\rm km\ s^{-1}$ \citep{crocker2012non,keeney2006does} to 1500 $\rm km\ s^{-1}$ \citep{strickland2009supernova}. Here, we use a moderate value $v_{\rm w}\approx 1000~\rm km\ s^{-1}$ for interacting galaxies since these galaxies may enter star-forming/starburst phase. In this case we have the advection time $t_{\rm ad}\simeq R_{\rm g}/v_{\rm w}\approx2.94\times10^6~{\rm yr} \left(\frac{v_{\rm w}}{1000\ \rm km\ s^{-1}}\right)^{-1}\left(\frac{R_{\rm g}}{3\ \rm kpc}\right).$}

{\yuan Inside the galaxy, the electron-positron injection spectrum can be modified due to additional injections via two-photon annihilation, $\gamma\gamma\rightarrow e^{-}e^+$, since the core region can be opaque to high-energy gamma-ray photons above a certain threshold energy $\varepsilon_{\gamma\gamma}^{\rm cut}$. In the pion decay scenario, the gamma-ray spectrum and the neutrino spectrum are correlated by $\varepsilon_\gamma^2\frac{dN_\gamma}{d\varepsilon_\gamma}=\frac{2}{3}\varepsilon_\nu^2\frac{dN_\nu}{d\varepsilon_\nu}|_{\varepsilon_\gamma=2\varepsilon_\nu}.$ From energy conservation, we may approximately relate the electron-positron injection rate to the gamma-ray production rate, and the former spectrum can be written as 
\begin{linenomath*}
\begin{equation}
\varepsilon^{2}\mathcal N_e^{\gamma\gamma}(\varepsilon_e)=2\varepsilon^{2}\frac{dN_\gamma}{d\varepsilon_\gamma}|_{\varepsilon_\gamma=2\varepsilon_e}=\frac{1}{3}\varepsilon^{2}\frac{dN_\nu}{d\varepsilon_\nu}|_{\varepsilon_\nu=\varepsilon_e},\ \varepsilon_{e}>\varepsilon_{\gamma\gamma}^{\rm cut}/2.
\label{eq:elec_pair}
\end{equation}
\end{linenomath*}

The total electron-positron injection spectrum is therefore the summation of Equations \ref{eq:elec_injection} and \ref{eq:elec_pair}, or equivalently we can introduce a modification factor $\chi(\varepsilon)=1+\exp(-\varepsilon_{\gamma\gamma}^{\rm cut}/2\varepsilon)$ to Equation \ref{eq:elec_injection}.
}

With these preparatory work, we can now derive the secondary electron-positron distributions and calculate the synchrotron and inverse Compton emissions. Considering the dynamic time $t_{\rm dyn}=R_{\rm g}/v_s$, we have the rate of lepton production
{\yuancc
\begin{linenomath*}
\begin{equation}
Q({\varepsilon},t)=\frac{\mathcal N_e(\varepsilon)\chi(\varepsilon)}{t_{\rm dyn}}\times\min\{1,e^{-\frac{t-t_{\rm dyn}}{t_{\rm esc}}}\},
\end{equation}
\end{linenomath*}
where the exponential factor describes the escape of CRs after the dynamical time scale and is obtained through the equation $\partial N/\partial t=-N/t_{\rm esc}$.} To get the electron distribution inside the galaxy, we solve the transport equation {\yuan of a simplified leaky-box model}
\begin{linenomath*}
\begin{equation}
\frac{\partial N_e}{\partial t}=Q(\varepsilon,t)-\frac{N_e}{t_{\rm esc}}+\frac{\partial}{\partial \varepsilon}[b(\varepsilon)N_e(\varepsilon,t)]
\label{eq:CR_transport}
\end{equation}
\end{linenomath*}

where $b(\varepsilon)$ is the electron energy loss rate due to synchrotron radiation, SSC/EIC and advection ($b_{\rm ad}\simeq \varepsilon/t_{\rm ad}$). In our calculations, we assume $Q$ and the diffusion coefficient $D_{\rm g}$ do not depend on the positions in the merging system. 

In the synchrotron limit $\gamma_e\gg 1$, the synchrotron radiation power in the frequency range $\omega$ to $\omega+d\omega$ by one electron with Lorentz factor $\gamma_e$ can be written in the well-known formula
\begin{linenomath*}
\begin{equation}
P_{\rm syn}(\omega,\gamma_e)d\omega=\frac{\sqrt{3}e^3B\sin\theta_p}{2\pi m_e c^2}F(X)d\omega
\label{eq:sync}
\end{equation}
\end{linenomath*}

where $\theta_p$ is the angle between the electron velocity and the magnetic field, which is assumed to be $\pi/2$ in our case,
\begin{linenomath*}
\[X=\frac{\omega}{\omega_c},\ \omega_c=\frac{3}{2}\gamma_e^2\frac{eB}{m_ec}.\]\end{linenomath*}

The function $F(X)$ is given by
\begin{linenomath*}
\[F(X)=X\int_{X}^\infty K_{5/3}(\xi)d\xi.\]\end{linenomath*}

Then, it is straightforward to write down the integrated radiation power
\begin{linenomath*}
\[b_{\rm syn}(\varepsilon)=\int P_{\rm syn}(\omega,\varepsilon/m_ec^2)d\omega.\]\end{linenomath*}

\begin{figure}\centering
\includegraphics[width=0.48\textwidth]{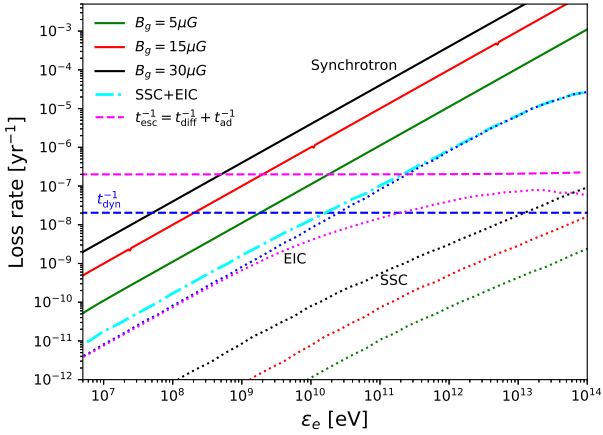}
\caption{Electron loss rates versus electron energy $\varepsilon_e$. Solid lines correspond to cooling rates due to synchrotron radiation in different magnetic fields, e.g. $5\mu {\rm G}$ (green), $15\mu {\rm G}$ (red) and $30\mu {\rm G}$ (black). The cyan dash-dotted line is the cooling rate of inverse Compton scattering (SSC+EIC). Blue and magenta dotted lines illustrate the contributions of CMB and EBL to the EIC cooling rate, while the black, red and green dotted lines are SSC cooling rates at the magnetic fields $5\mu {\rm G}$ (green), $15\mu {\rm G}$ (red) and $30\mu {\rm G}$ (cyan), respectively. Magenta and blue dashed lines are the escape rate and the reciprocal of dynamic time, respectively.}
\label{fig:sync_cooling}
\end{figure}

\begin{figure}\centering
\includegraphics[width=0.48\textwidth]{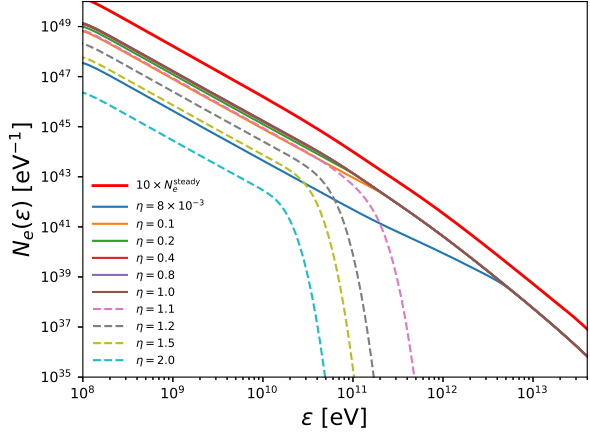}
\caption{Secondary electron-positron spectra at different times assuming the magnetic field $B=5~\mu {\rm G}$. The parameter $\eta=t/t_{\rm dyn}$ represents the time of electron-positron injection. Thin lines are numerical solutions to the CR transport equation while the thick red line is the analytical steady-state solution $N_e^{\rm steady}$. To separate $N_e^{\rm steady}$ from numerical solutions, we multiply $N_e^{\rm steady}$ by a factor of 10. }
\label{fig:elec_spec}
\end{figure}

It is useful to define the synchrotron cooling time
\begin{linenomath*}
\begin{equation}
t_{\rm syn}(\varepsilon)=\frac{\varepsilon}{b_{\rm syn}(\varepsilon)}.
\end{equation}\end{linenomath*}

While SSC and EIC also play a role in electron-positron cooling, we will show later that these processes are subdominant comparing to synchrotron cooling. Now with the preparations on synchrotron radiation, we are able to solve the  kinetic equation. One special solution to the differential equation is the steady state solution ($\partial N_e/\partial t=0$),
\begin{linenomath*}
\begin{equation}
N_e^{\rm steady}=Q(\varepsilon,t)\left(\frac{1}{t_{\rm esc}}+\frac{1}{t_{\rm syn}}\right)^{-1}
\label{eq:steady_sol}
\end{equation}\end{linenomath*}

To verify this expression, it is worthwhile to solve the time evolution of electron-positron spectra numerically. For illustration purposes, we assume $M_{\rm g}=10^9~M_\odot$, $v_s=100~\rm km\ s^{-1}$, $R_{\rm g}=5~\rm kpc$ and $\varepsilon_{\gamma\gamma}^{\rm cut}=1~\rm TeV$. Figure \ref{fig:sync_cooling} shows the synchrotron cooling rate ($t_{\rm syn}^{-1}$; solid lines) as functions of lepton energy for different galactic magnetic fields as well as the escape rate ($t_{\rm esc}^{-1}$) and the reciprocal of dynamic time ($t_{\rm dyn}^{-1}$; dashed lines). As the magnetic field becomes stronger, the synchrotron cooling tends to be faster since $P(\omega,\gamma_e)$ increases. Using the finite difference method, the time evolution of pair spectra for the magnetic field $B=5~\mu {\rm G}$ is shown in the Figure \ref{fig:elec_spec}, where we use the parameter $\eta=t/t_{\rm dyn}$ to label the stages of pair injection. The thick red solid line corresponds to the steady electron distribution given by the Equation \ref{eq:steady_sol}. The theoretical steady distribution almost coincide with the numerical steady solutions. To show this, we multiply the theoretical solution $N_e^{\rm steady}$ by a factor of ten to separate these curves. Figure \ref{fig:elec_spec} also illustrates the evolution of the cumulative number of electron inside the core region. From this figure, we conclude that the electron injection enters the steady phase when $\eta\gtrsim0.2$.

Inverse Compton scattering between high-energy electron-positron pairs and external CMB/SL/EBL photons (denoted by EIC) as well as SSC may become more pronouncing in lepton cooling process when the electron-positron spectrum becomes harder. Here we formulate the SSC/EIC power per unit comoving volume as \citep[e.g.,][]{murase2011implications},

\begin{linenomath*}
\begin{equation}
	E\frac{dN_{x}}{dEdt}=\int d\gamma_e \frac{dN_e}{d\gamma_e}\int d\varepsilon_\gamma \left(\frac{dn_{\gamma}}{d\varepsilon_\gamma}\right)_x E\left\langle c'\frac{d\sigma_{IC}}{dE}\right\rangle
\end{equation}\end{linenomath*}

where $x=$ SSC or EIC, and the differential cross section is \citep{blumenthal1970bremsstrahlung}:
\begin{linenomath*}
\begin{equation}
\begin{split}
\left\langle \frac{d\sigma_{IC}}{dE}c'\right\rangle=&\frac{3}{4}\sigma_Tc\frac{1}{\gamma_e^2\varepsilon_\gamma}\times\\
		&\left[1+v-2v^2+\frac{v^2w^2(1-v)}{2(1+vw)}+2v\ln v \right].
		\end{split}
		\end{equation}\end{linenomath*}

In the expression of the cross section, $\sigma_T$ is the Thomson cross section, $v=\frac{E}{4\varepsilon\gamma_e^2(1-\xi)}$, $\xi=E/(\gamma_e m_ec^2)$ and $w=\frac{4\varepsilon_\gamma\gamma_e}{m_ec^2}$. For SSC, $dn_\gamma/d\varepsilon_\gamma$ corresponds to the photon spectrum of synchrotron emission and it can be written as
\begin{linenomath*}
\begin{equation}
\varepsilon_\gamma\left(\frac{dn_\gamma}{d\varepsilon_\gamma}\right)_{\rm SSC}=\frac{1}{{2}R_{\rm g}^2ch}\int P_{\rm syn}\left(\varepsilon_\gamma/\hbar, \frac{\varepsilon_e}{m_ec^2}\right)N_e(\varepsilon_e)d\varepsilon_e
\label{eq:SSC_photon}
\end{equation}\end{linenomath*}

\begin{figure*}\centering
\includegraphics[width=0.49\textwidth]{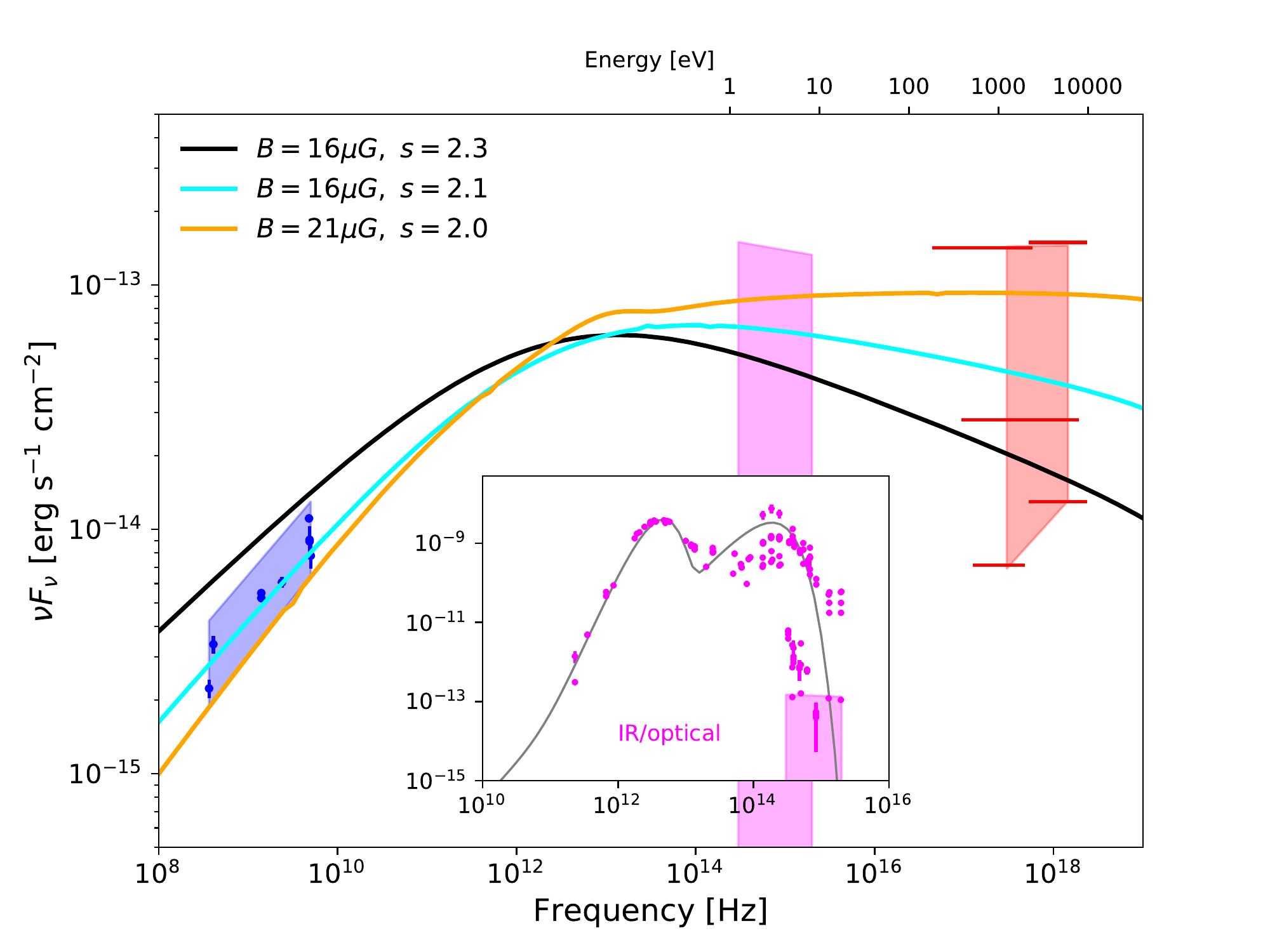}
\includegraphics[width=0.49\textwidth]{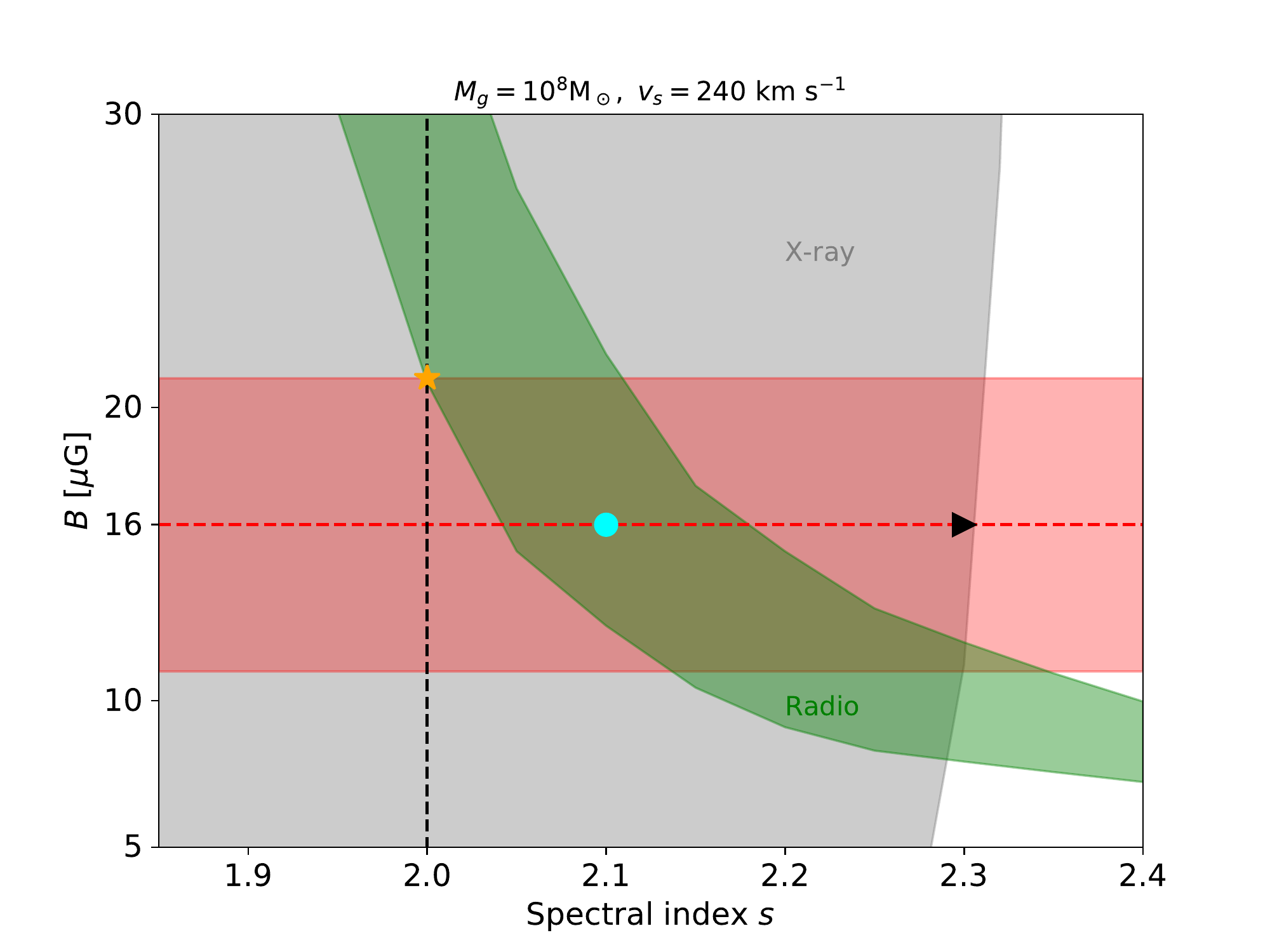}
\caption{Left panel shows the spectral energy distribution for NGC 660, extending from the radio band to the X-ray regime.  Blue points are radio fluxes at various frequencies and the red points are X-ray data in the energy range $0.2-10~\rm keV$. Observations from microwave to UV are illustrated as magenta points in the inset. The gray line is the Planck-function approximation to the IR/optical data. The bulk of the microwave, IR and optical spectrum is due to starlight and dust re-radiation. The fitting areas of radio, UV and X-ray data are shown as the blue, magenta and red areas, respectively. The black, cyan and orange lines are spectra that correspond to the black wedge, cyan circle and orange star in the right panel. In the right panel, the gray and green areas are X-ray and radio constraints on $s-B$ plane. The red area shows the constraints on the magnetic from previous polarization studies, $16\pm5\ \mu\rm G$.}
\label{fig:spec}
\end{figure*}

{\yuan The intergalactic starlight photon density can be estimated by using the IR/optical spectral energy density (SED; see the inset of the left panel of Figure \ref{fig:spec}), e.g. $\varepsilon_\gamma(dn/{d\varepsilon_\gamma})_{\rm SL}\sim {2}d_L^2F_{\nu,\rm SL}/(R_g^2ch)$, where $d_L$ is the luminosity distance of the galaxy. In this paper, we use two modified Planck functions to approximate the left and right bulks of the IR/optical data,
\begin{linenomath*}
\begin{equation}
    F_{\nu,\rm SL}(\nu)=\sum_{i=L,R}A_i \left(\frac{h\nu}{1\rm eV}\right)^{\zeta_i}\frac{1}{\exp(\frac{h\nu}{\varepsilon_i})-1}.
    \label{eq:starlight}
\end{equation}\end{linenomath*}

}As for EIC, $(dn_\gamma/d\varepsilon_\gamma)_{\rm EIC}$ is given by the summation of CMB black body spectrum, $(dn/{d\varepsilon_\gamma})_{\rm SL}$ and the EBL photon density spectrum provided by {\yuan{"model C" from}} \cite{finke2010modeling}.

Like the synchrotron radiation, we can define the cooling time for SSC and EIC, 
\begin{linenomath*}
\begin{equation}
t_x(\varepsilon_e)={\varepsilon_e}\left[\int dE\int d\varepsilon_\gamma \left(\frac{dn_{\gamma}}{d\varepsilon_\gamma}\right)_x E\left\langle c'\frac{d\sigma_{IC}}{dE}\right\rangle\right]_{\gamma_e=\frac{\varepsilon_e}{m_ec^2}}^{-1},
\end{equation}\end{linenomath*}

The cyan dash-dotted line in Figure \ref{fig:sync_cooling} shows the combined cooling rate $t_{\rm IC}^{-1}=t_{\rm SSC}^{-1}+t_{\rm EIC}^{-1}$ as a function of electron energy. Figure \ref{fig:sync_cooling} illustrates also the components of the total IC cooling rate, e.g. CMB (blue dotted line), EBL (magenta dotted line) and SSC at the magnetic fields $30~\mu {\rm G}$ (black dotted line), $15~\mu {\rm G}$ (red dotted line) and $5~\mu {\rm G}$ (green dotted line). {\yuancc The flattening of the EIC loss rate is due to the Klein-Nishina regimes as the electron Lorentz factor increases.} From this figure, we find that the cooling process is dominated by synchrotron radiation and the cooling due to EIC is predominant comparing to SSC. Hence, in the following section where the application to NGC 660 is discussed, we only consider $t_{\rm syn}$ in the CR transport equation (Equation \ref{eq:CR_transport}). In general, for a power-law electron distribution, the SSC cooling rate should have the same slope. However, in Figure \ref{fig:sync_cooling}, the physical cause of the slight slowing down of the growth of the SSC cooling rate is that the steady-state electron spectrum becomes steeper due to synchrotron cooling (see the red line in Figure \ref{fig:elec_spec}) and this can influence the synchrotron photon density spectrum through Equation \ref{eq:SSC_photon}. With the equations above, we can write down the equations for synchrotron and SSC/EIC fluxes
\begin{linenomath*}
\begin{equation}\begin{split}
&F_{\nu}^{\rm syn}=\frac{1}{4\pi d_L^2}\int 2\pi\cdot P_{\rm syn}\left(2\pi\nu,\frac{\varepsilon_e}{m_ec^2}\right)N(\varepsilon_e)d\varepsilon_e\\
&F_{\nu}^{x}=\frac{h}{4\pi d_L^2}\left[E\frac{dN_{x}}{dEdt}\right]_{E=h\nu},\ x=\rm SSC\ or\ EIC, 
\label{eq:fluxes}
\end{split}
\end{equation}\end{linenomath*}

where the coefficient $2\pi$ and Planck constant $h$ come from $|d\omega/d\nu|$ and $|dE/d\nu|$, respectively. In general, we need to keep in mind that inverse Compton (or more especially SSC) emission can be significant at some frequency even when the magnetic field is strong and the core region is more compact such that the synchrotron photon field is more intense. We will show later that SSC and EIC can also be important for NGC 3256.

\section{Radio and X-ray constraints on $M_{\rm g}$ and $v_s$}
\label{sec:application}

With the above, we are able to calculate the synchrotron and SSC/EIC fluxes. The spectrum of synchrotron radiation extends broadly from radio band to X-ray regime while SSC/EIC may become important from optical band to X-ray band. In this section we investigate the possibility of explaining the radio and X-ray observations simultaneously using the formalism presented in \S\ref{sec:elec_input}. Since in our model the physical state of the core region of merging galaxies is determined by five parameters: the radius $R_{\rm g}$, the magnetic field $B$, the gas mass $M_{\rm g}$, the shock velocity $v_s$ and the time parameter $\eta=t/t_{\rm dyn}$, our model provides one useful method to study the dynamics of galaxy mergers. 
In this section, we present an application to the interacting system NGC 660 and show that our model can be used to reproduce the radio and X-ray observation. In addition, we find that $M_{\rm g}$ and $v_s$ in the core region of NGC 660 can be constrained under appropriate assumptions. To show that our model can be used widely to general galaxy merging systems, we consider also the galaxy NGC 3256. From Figure \ref{fig:elec_spec}, we find that the interacting system can be approximately treated as a steady state. Hence, to simplify the constraint, we employ a steady state solution to approximate the secondary electron-positron distribution throughout the paper.

\begin{figure*}\centering
\includegraphics[width=0.328\textwidth]{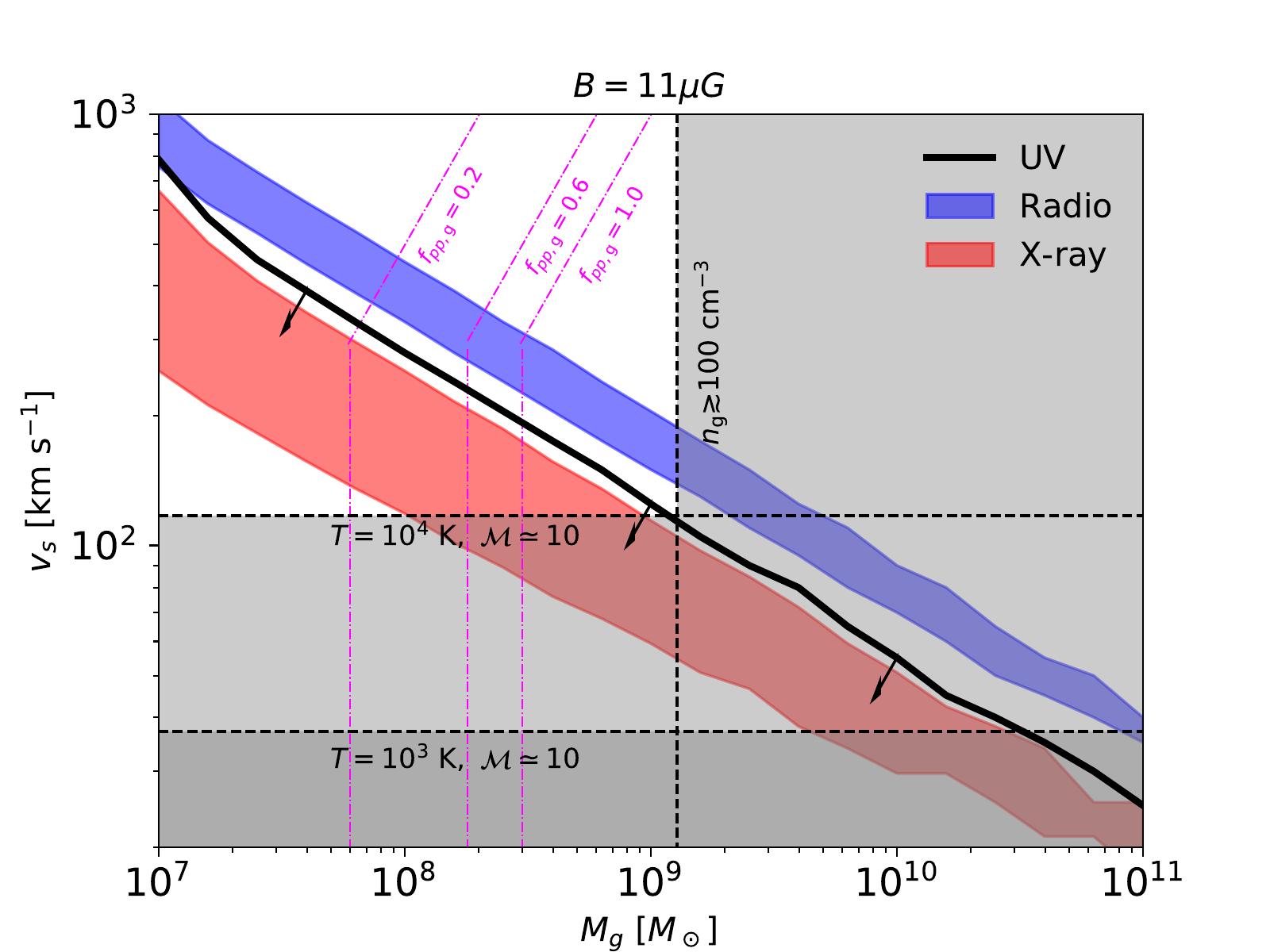}
\includegraphics[width=0.328\textwidth]{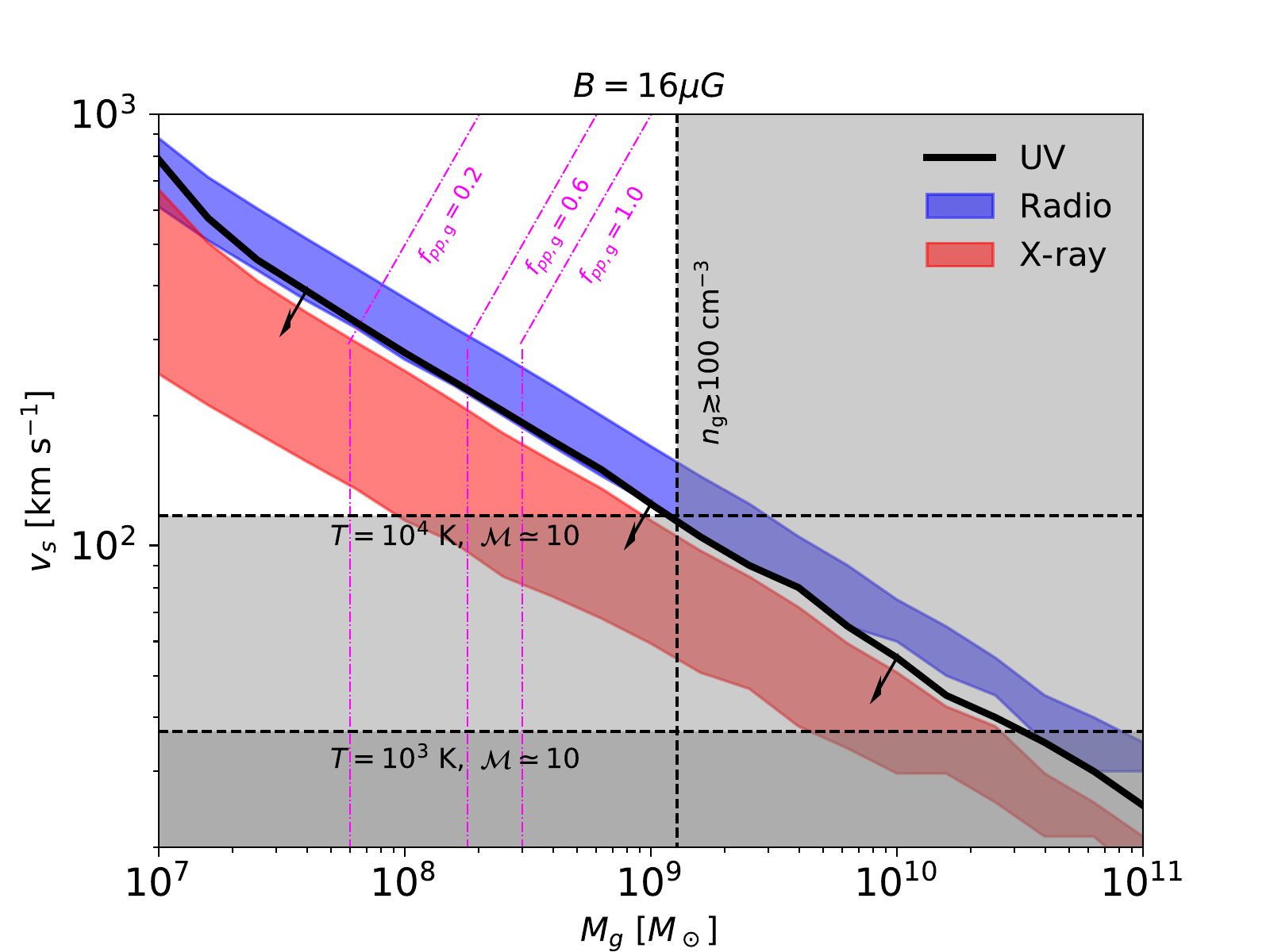}
\includegraphics[width=0.328\textwidth]{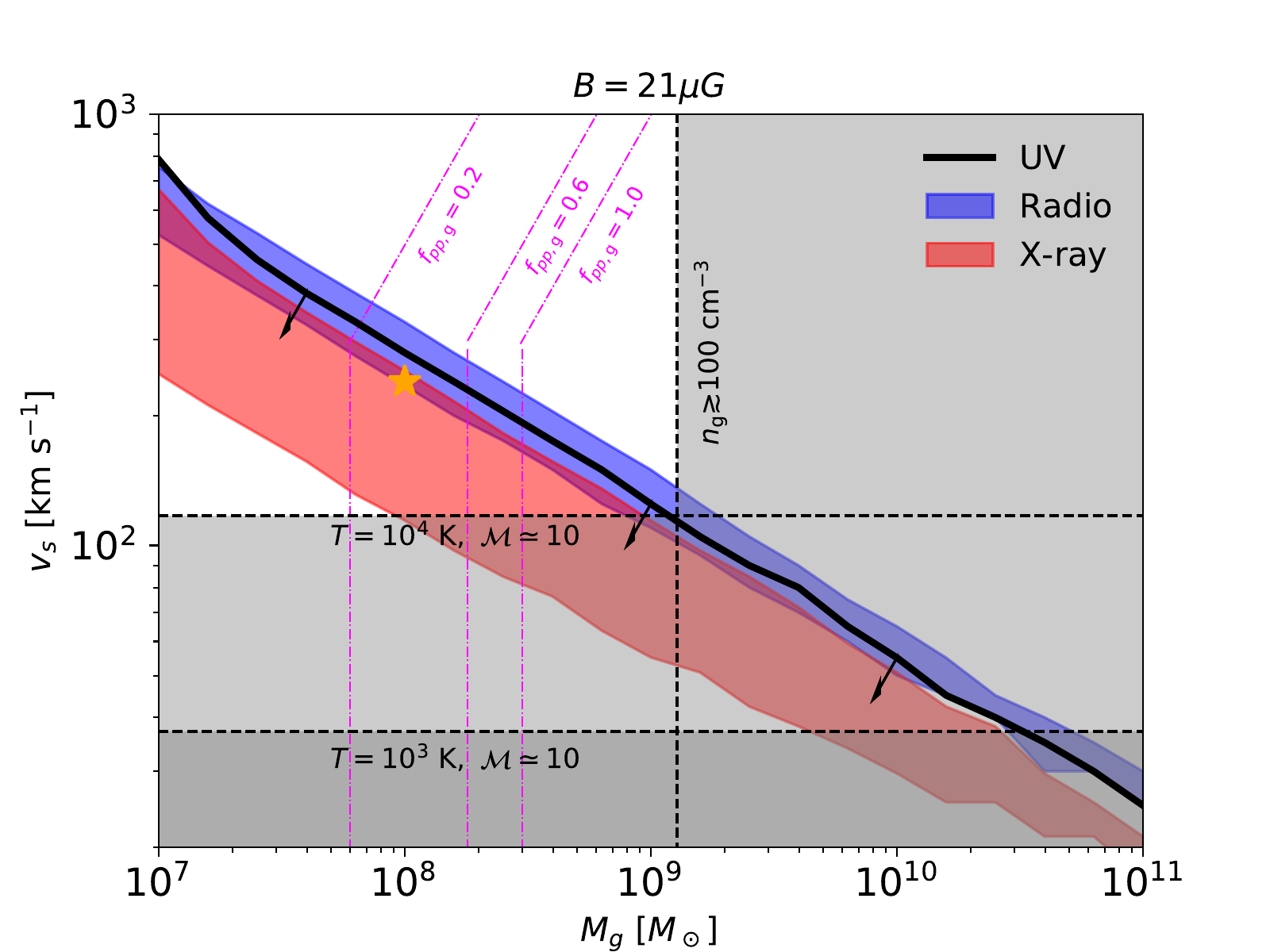}
\caption{Constraints on $M_{\rm g}-v_s$ plane from radio, UV and X-ray tolerance areas. From left to right, magnetic fields are assumed to be $B=11~\mu {\rm G},\ 16~\mu {\rm G}$ and $21~\mu {\rm G}$. In each figure, blue and red areas correspond to the radio and X-ray constraints and the black line shows the upper boundary under the UV constraint. {\yuanccs The vertical dashed line and gray area show the constraints from the core region gas density $n_{\rm g}\lesssim100\rm \ cm^{-3}$, whereas the horizontal dashed lines and gray area correspond to the strong shock requirements ($\mathcal M\simeq10$) for the temperature $10^{4}$ K and $10^{3}$ K. The magenta dash-dotted contours correspond to different $pp$ optical depth $f_{pp,\rm g}$.} The orange star in the overlapping region labels the test case: $B=21\ \mu{\rm G},\ v_s={\yuancc 240}\ {\rm km\ s^{-1}},\ M_g=10^8\ \rm M_\odot$. }
\label{fig:constraints}
\end{figure*}

\subsection{NGC 660}

NGC 660 is usually believed as a galaxy formed by the collision and merger of two galaxies. The distance to us is $d_L\sim12.3\rm Mpc$ and the HI extent is $47~\rm kpc$. Radio maps by VLA  reveal a smooth core region \citep{condon1982strong}. {\yuancc \cite{fraternalif2004further} showed that the de-convoluted angular size of the radio and X-ray emitting region is less than 10 arcsec or equivalently the radius $R_{\rm g}\lesssim 0.5$ kpc. Hence, in our calculations, we use $R_{\rm g}\simeq0.5$ kpc as the fiducial radius of the core region.} In addition, \cite{drzazga2011magnetic} studied the magnetic fields using VLA data in 16 interacting galaxies and they find that the average magnetic field of NGC 660 is $16\pm5~\mu {\rm G}$. In the X-ray regime, the data from Chandra telescope gives the X-ray flux $1.24^{+0.37}_{-0.54}\times10^{-13}~\rm erg\ cm^{-3}\ s^{-1}$ \citep{argo2015new} in the range $0.5-10~\rm keV$. In mid-2013, a radio outburst was observed using e-MERLIN and after the outburst the X-ray flux also increased to $1.85^{+0.19}_{-0.16}\times10^{-13}~\rm erg\ cm^{-3}\ s^{-1}$. The origin of the outburst was investigated in mid-2013 and it might be produced by AGN activities in the galaxy center. In our work, we focus on the emissions from the smooth core region, therefore we use the data recorded before the outburst. Above all, with the magnetic field estimated by \cite{drzazga2011magnetic}, the parameters left to be determined are $M_{\rm g}$ and $v_s$. 

The left panel of Figure \ref{fig:spec} shows the spectral energy distribution for NGC 660 from radio band to X-ray band.  Blue points are radio fluxes at 365 MHz \citep{douglas1996texas}, 408 MHz \citep{large1981molonglo}, 1.4 GHz \citep{condon2002radio,condon1998nrao} 2.38 GHz \citep{dressel1978arecibo}, 4.78 GHz \citep{bennett1986green}, 4.85 GHz \citep{becker1991new,gregory199187gb} and 5 GHz \citep{sramek19755}. The red points are X-ray data before the radio burst in the energy range $0.2-10~\rm keV$, which are provided by Chandra \citep{fraternali2004further,liu2010chandra}, XMM-Newton \citep{brightman2011xmm} and ROSAT \citep{white2000wgacat}. {\yuan Since this source was observed with short exposure times, the photon count rates were converted to the X-ray fluxes by assuming a spectral index in the energy range for each red bar in this figure. More details on the data reductions can be found in the corresponding references.} {\yuancc In our model, the synchrotron spectrum can reproduce the slope of radio spectra, which is the primary motivation of our work. However as for the X-ray data, the slope is quite uncertain and depends on different observations and models. Therefore in the X-ray band we attempt to explain integrated fluxes from different observations in the energy range $0.2-10\ \rm keV$.} The broadband observations from microwaves to UV are shows as magenta points in the inset.\footnote{A full list of references can be found in the page \href{http://ned.ipac.caltech.edu/cgi-bin/objsearch?search_type=Obj_id&objid=4674&objname=1&img_stamp=YES&hconst=73.0&omegam=0.27&omegav=0.73&corr_z=1}{NED:INDEX NGC 660}}. {\yuan The gray line shows the approximation to the IR/optical data using Equation \ref{eq:starlight}, with the parameters $A_L=5.15\times10^{10}\ {\rm Jy},\ \zeta_L=3.9,\ \varepsilon_L=0.004\ {\rm eV}; A_R=3.44\ {\rm Jy},\ \zeta_R=1.8,\ \varepsilon_L=0.3\ {\rm eV}$.} To measure the consistency between synchrotron spectrum and the observations, we set three fitting areas, as shown in the left panel of Figure \ref{fig:spec}.  The blue and red areas correspond to the error tolerances of radio and X-ray data respectively. As for the microwave, infrared and UV data points, we need to keep in mind that the dust in the galaxy and star forming activities may dominate the emissions in these bands. Hence we assume the secondary radiation in the shock region merely contributes to the background and use the UV data as the upper limit in our model (see the magenta area). One vexing problem of the UV limit is that the dust absorption in the host galaxy cannot be neglected and the photometry correction is model dependent. Hence, in our calculation, we use the UV limit just as a reference. 

{\yuanccs{NGC 660 has been identified as a star-forming/starburst galaxy \citep{van1995polar}, which provides one complementary constraint on the gas mass once the radius $R_{\rm g}$ is specified. The gas density in starburst galaxies can be up to $n_{\rm g}\simeq100\rm\ cm^{-2}$ and thus we conclude that the gas mass in the core region satisfies $M_{\rm g}\lesssim\frac{4\pi}{3}\mu m_p n_{\rm g} R_{\rm g}^3$}, where $\mu\simeq1.24$ is the mean molecular weight. The vertical dashed lines and the gray areas in Figure \ref{fig:constraints} illustrate the gas density constraint. Another caveat is that a strong shock with the Mach number $\mathcal M\gtrsim10$ is required to produce a power-law electron spectrum with index $s\simeq2(\mathcal M^2+1)/(\mathcal M^2-1)\simeq2.$ Observations reveal that NGC 660 has the dust temperature and kinetic temperature around $40$ K and $200$ K \citep{van1995polar,mangum2013ammonia}, respectively. Here, we use $T\simeq10^4\rm \ K$ as an optimized value since the core region may contain warm gas and evaluate the lower limit of the shock velocity $v_s\gtrsim \mathcal M \sqrt{\frac{\gamma k_B T}{\mu m_p}}$ which is shown as the upper horizontal dashed lines and gray areas in Figure \ref{fig:constraints}. For illustration purpose, we show also the constraint obtained by assuming a relatively lower temperature $T=10^3$ K (the lower dashed lines). Meanwhile, we include the contours of $pp$ optical depth $f_{pp,\rm g}$ in the $v_s-M_\odot$ plane (magenta dash-dotted lines in Figure \ref{fig:constraints}). As we can see, $pp$ interactions are more efficient in a region with large gas mass and low shock velocity $v_s$ due to the higher gas density $n_{\rm g}$ and longer collision time. When $v_s$ decreases to one critical value, which is determined by $t_{\rm dyn}=t_{\rm esc}$, the particle escape dominates the interaction time. Therefore, the oblique lines become vertical.}

Considering the uncertainty of magnetic field, we select $B=11~\mu {\rm G},\ 16~\mu {\rm G}$ and $21~\mu {\rm G}$ as three fiducial values. Figure \ref{fig:constraints} shows the constraints on $M_{\rm g}-v_s$ plane from the radio, UV and the X-ray error tolerances (see the blue, magenta and red areas in Figure \ref{fig:spec}).  From these figures, we find that the permissible areas in the $M_{\rm g}-v_s$ plane overlap only at higher magnetic fields, which means that to fit the radio, UV and X-ray data simultaneously, a stronger magnetic field is favored. 
This conclusion is also consistent with the {\yuan orange line in the left panel of Figure \ref{fig:spec}, which shows the flux predicted by our model for the test point, the orange star ($B=21\ \mu{\rm G},\ M_g=10^8\ {\rm M_\odot},\ v_s={\yuancc 240}\ \rm km\ s^{-1}$), in the overlapping region of Figure \ref{fig:constraints}. Meanwhile, we find that the contributions from SSC and EIC are subdominant comparing with synchrotron emissions in the case of NGC 660.} 
For a lower magnetic field, the tension between radio data and X-ray data is inevitable. To fit the radio data, the synchrotron spectrum will overshoot X-ray flux and UV upper limit. On the other hand, to alleviate the tension, we need to make the synchrotron spectrum higher in the radio regime while keeping the X-ray flux unchanged. This can be achieved by increasing the magnetic field, since the synchrotron spectra converge at high energy band (e.g. X-ray) even if we increase the magnetic field. We provide one brief proof here.  From Figure \ref{fig:sync_cooling}, we see that synchrotron cooling dominate the electron spectrum ($t_{\rm syn}^{-1}\gg t_{\rm esc}^{-1}$) when the electron energy is high, which means $N_e^{\rm steady}\simeq Q(\varepsilon_e,t)t_{\rm syn}=\frac{\varepsilon_e Q(\varepsilon_e,t)}{b_{\rm syn}(\varepsilon_e)}$. Combining $N_e^{\rm steady}$ with Equations \ref{eq:sync} and \ref{eq:fluxes}, we obtain 
\begin{linenomath*}
\begin{equation}\begin{split}
F_{\nu}^{\rm syn}&\propto \int \varepsilon_e Q(\varepsilon_e,t)\frac{P(\omega,\varepsilon_e)}{\int P(\omega',\varepsilon_e)d\omega'}d\varepsilon_e\\
&\propto  \int \varepsilon_e Q(\varepsilon_e,t)\frac{F(X)}{\int F(X')d\omega'}d\varepsilon_e. 
\end{split}
\label{eq:sync_magnetic}
\end{equation}\end{linenomath*}
At high energy limit, the function $F(X)$ has the asymptotic form $F(X)\simeq \sqrt{2\pi X}e^{-X}$ and the flux no longer depends on the magnetic field. {\yuancc A more physical interpretation is that once $B$ is high enough, the energy of electrons is radiated away through synchrotron fast cooling. In this case, the flux only depends on the electron injection rate. Meanwhile, it's easy to see that the flux will increase as $B$ increases in a lower energy band (e.g. radio regime) since electrons lose more energy in a stronger magnetic field}. Above all, {for a flat CR spectrum with the spectral index $s\sim2$}, a higher magnetic field will keep the X-ray flux unchanged with increasing the radio flux and therefore can be used to fit the radio and X-ray data simultaneously.

{\yuan This simple single-zone model meets difficulty explaining the radio and X-ray observations at the same time with a relatively lower $B$. This motivates us to exploit the chance of improving the fitting by varying the CR spectral index $s$ in the range 1.8-2.4. As $s$ deviates from 2.0, the normalization coefficient in Equation \ref{eq:elec_injection} changes to $(\varepsilon_{\rm max}^{2-s}-\varepsilon_{\rm min}^{2-s})/(2-s)$ and a correction factor $\varepsilon^{2-s}$ should be applied to the electron spectrum. To demonstrate the impact of $s$ and $B$ on the fitting, we select and fix the gas mass and shock velocity to be $10^8\ {\rm M_\odot}$ and ${\yuancc 240}\ \rm km\ s^{-1}$, the orange star in the overlapping region in Figure \ref{fig:constraints}. The right panel of Figure \ref{fig:spec} shows the constraints in the $s-B$ plane from polarization studies (red area), radio (green area) and X-ray (gray area) observations. Firstly, we find that magnetic field almost does not influence the X-ray results, which is consistent with the previous analysis. There exist a cut off around $s=2.35$, beyond which the X-ray flux could be too low to explain the observations. Secondly, as the index $s$ increases, the electron spectrum becomes steeper, or on other words, more low-energy electrons are injected. Consequently, radio flux got flattened while X-ray flux steepened. Therefore, a low magnetic field is required to counteract radio flux increase and as a result we expect the green area for radio constraint. One straightforward conclusion we can make from this figure is that, a relative larger spectral index can be used to reproduce the radio and X-ray data simultaneously, e.g. the  parallelogram region formed by the green and red areas. To show that explicitly, we select three representative points in the $s-B$ plane, e.g. orange star $(s=2.0,\ B=21\ {\mu\rm G})$, cyan circle ($s={\yuancc 2.1},\ B=16\ \mu\rm G$) and black wedge $(s=2.3,\ B=16\ \mu\rm G)$. The corresponding X-ray and radio fluxes are shown in the left panel of Figure \ref{fig:spec}. Obviously, from this figure, a moderately larger $s$ in the range $\sim2.1-2.2$ with the optimized magnetic field $B=16~\mu\rm G$ can provide a good fitting. These indices are also consistent with the observations of starburst galaxies such as M82 and NGC 253.}

From the discussions above, we showed that our one-zone model can be used to explain the radio, UV and X-ray observations of the NGC 660 core region. Given our model is correct, one can constrain the gas mass $M_{\rm g}$, magnetic field $B$, CR spectral index $s$ and collision velocity $v_s$ in that region. 

\begin{figure*}\centering
\includegraphics[width=0.49\textwidth]{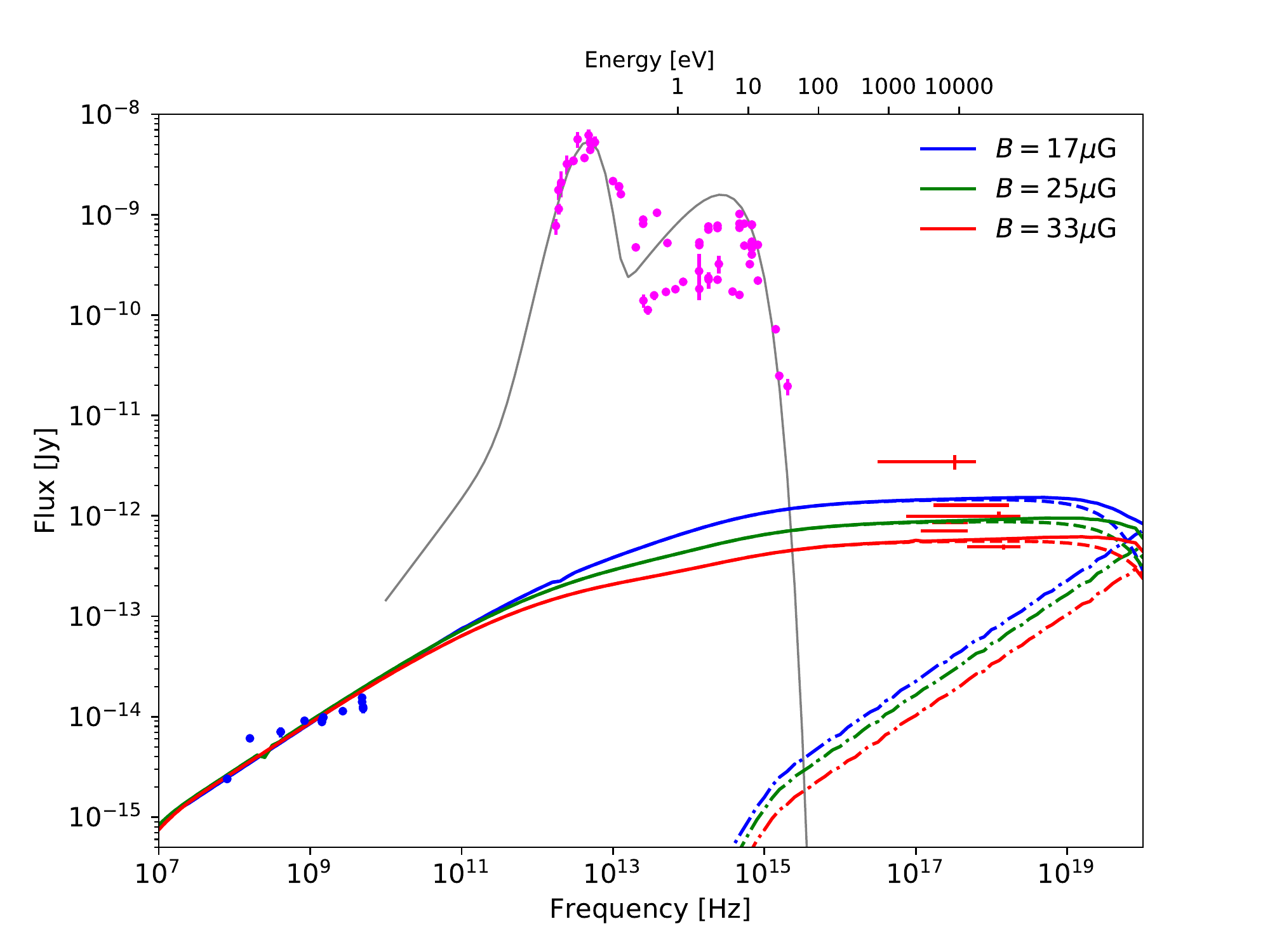}
\includegraphics[width=0.49\textwidth]{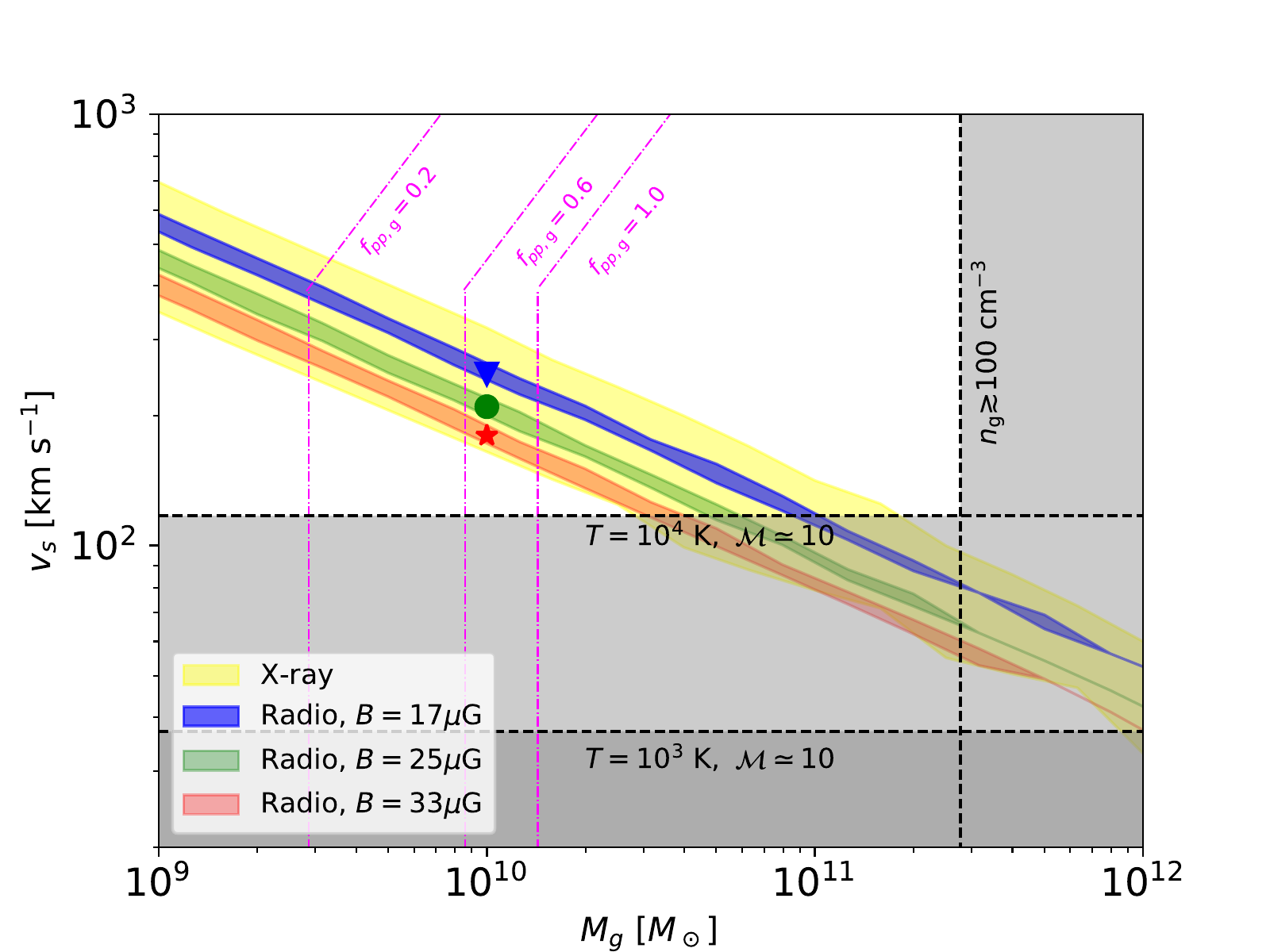}
\caption{Left panel: The spectral energy distribution for NGC 3256. Blue and red points are radio and X-ray fluxes, respectively. The observations from the infrared band to the UV band, which are mainly attributed to dust and starlight, are shown as magenta points. The blue, green and red lines are best-fitting spectra obtained from three selected points in the right panel for different magnetic fields. The dashed and dash-dotted lines correspond to the synchrotron and IC components. The right panel shows the X-ray (yellow area) and radio constraints for the magnetic fields $17\ \mu\rm G$ (blue area), $25\ \mu\rm G$ (green area) and $33\ \mu\rm G$ (red area). The gray areas, black dashed lines and magenta dash-dotted lines have the same meaning with Figure \ref{fig:constraints}.}
\label{fig:NGC 3256}
\end{figure*}

\subsection{NGC 3256}

NGC 3256 is also a galaxy formed by the collision of two galaxies and the redshift of NGC 3256 is $z\approx0.009364$ \citep{meyer2004hipass}. In a $\Lambda$CDM universe with $\rm \Omega_m=0.286$ and $H_0=69.6\rm\ km\ s^{-1}\ Mpc^{-1} $, the luminosity distance to us is $d_L=40.6\ \rm Mpc$.  It provides a nearby template for studying the properties of merging galaxies. Nearly infrared observations \citep{skrutskie2006two} reveal that the major axis and minor axis sizes are $a=$1.277 arcmin and $b=$1.251 arcmin respectively. In our calculation, we assume an equivalent angular size $\theta_{\rm g}=\sqrt{ab}=1.264$ arcmin and the corresponding radius $R=14.92\rm\ kpc$. {\yuan However, instead of using the galaxy radius, we focus on the core/nucleus region where collisions occur. \cite{laine2003hubble} investigated the morphology of many merging galaxies including NGC 3256 using Hubble Space Telescope WFPC2 camera and the radius of the core region of NGC 3256 is approximately $3\ \rm kpc$. In the following calculations, we adopt $R_g=3~\rm kpc$.} Like NGC 660, \cite{drzazga2011magnetic} also provided the average magnetic field for NGC 3256, which is $25\pm8\ \mu {\rm G}$. Therefore, in this section we us $17\ \mu {\rm G}$, $25\ \mu {\rm G}$ and $33\ \mu {\rm G}$ as three fiducial values of the magnetic field. In the $0.3-10~\rm\ keV$ band, NGC 3256 has been observed by ASCA Medium Sensitivity Survey \citep{ueda2001asca}, XMM-Newton \citep{pereira2011x,jenkins2004xmm} and ROSAT \citep{brinkmann1994x}. As for the radio band, we use the data from broad-band observations in the frequency range 80 MHz to 5.0 GHz \citep{slee1995radio,large1981molonglo,condon19961,wright1994parkes,whiteoak1970observations}. Blue and red points in left panel of Figure \ref{fig:NGC 3256} show the radio and X-ray fluxes respectively. In this figure, we also plot the fluxes from infrared to UV bands as magenta points \footnote{A full list of references can be found in the page \href{http://ned.ipac.caltech.edu/cgi-bin/objsearch?objname=NGC 3256&extend=no&hconst=73&omegam=0.27&omegav=0.73&corr_z=1&out_csys=Equatorial&out_equinox=J2000.0&obj_sort=RA+or+Longitude&of=pre_text&zv_breaker=30000.0&list_limit=5&img_stamp=YES}{NED:INDEX for NGC 3256}}.  {\yuan The gray line in this figure is our approximation to the IR/optical data with the parameters $A_L=6.87\times10^{10}\ {\rm Jy},\ \zeta_L=3.9,\ \varepsilon_L=0.004\ {\rm eV};\ A_R=2.06\ {\rm Jy},\ \zeta_R=1.0,\ \varepsilon_L=0.7\ {\rm eV}$.}

Using the same procedure for NGC 660, we attempt to reproduce the observations of  NGC 3256. We find that we can fit the radio and X-ray data simultaneously {in the whole magnetic field range $17\ \mu{\rm G}-33\ \mu\rm G$ \yuan by using a simple CR spectral index $s=2$. The right panel illustrates the constraints from X-ray and radio observations. The X-ray constraint (yellow area) remains unchanged as consequence that the flux in X-ray band is not sensitive to the magnetic field. Radio constraints at $17\ \mu{\rm G},\ 25\ \mu{\rm G}$ and $33\ \mu{\rm G}$ are shown as blue, green and red areas.  {\yuanccs Like Figure \ref{fig:constraints}, the gray areas and black dashed lines correspond to the gas density and strong shock constraints.} Using the magnetic field given by polarization studies, our model can explain a significant fraction of X-ray flux. Left panel shows the spectra of three test points in the right panel, e.g. blue wedge ($17\ \mu{\rm G,\ 10^{10}\ {\rm M_\odot}},\ 250\ \rm km\ s^{-1}$), green circle ($25\ \mu{\rm G,\ 10^{10}\ {\rm M_\odot}},\ 210\ \rm km\ s^{-1}$) and red star ($33\ \mu{\rm G,\ 10^{10}\ {\rm M_\odot}},\ 180\ \rm km\ s^{-1}$). As anticipated, to fit the radio data, a stronger magnetic field implies a lower X-ray flux (see the red line). As for NGC 3256, since the radius of the nucleus is smaller and the starlight photon density is proportional to $(d_L/R_g)^2$, the starlight contribution to EIC is more significant than NGC 660. Meanwhile, considering that strong magnetic field can also boost SSC, in this case inverse Compton scattering is no longer negligible. The dashed lines and dash-dotted lines in the left panel of Figure \ref{fig:NGC 3256} show the synchrotron and IC contributions for various magnetic fields.

Above all, our simple one-zone model with $s\sim2$ can be used to explain the radio and a large fraction of X-ray observation and the constraint is in good agreement with previous magnetic studies.}

\section{Summary and discussion}
\label{sec:summary}
In this paper, we have investigated the synchrotron and SSC/EIC emissions from secondary electron-positron pairs in merging galaxies and found that these emissions can be used to reproduce the radio and X-ray observations of such systems, as calculated in detail for two of the best-studied galaxies formed by galaxy mergers, NGC 660 and NGC 3256. Combining the magnetic field in the core regions measured through polarization analyses, we showed that our model can be used to constrain the gas mass $M_{\rm g}$ and shock velocity $v_s$ under a steady-state approximation for the electron-positron  distribution. For NGC 660, in order to alleviate the tensions between the radio and X-ray constraints, a higher magnetic field $16~\mu {\rm G}\lesssim B\lesssim 21~\mu {\rm G}$ is required, which is consistent with the uncertainty of the magnetic field given by \cite{drzazga2011magnetic}. Utilizing $16~\mu {\rm G}\lesssim B\lesssim 21~\mu {\rm G}$ as the fiducial range of magnetic field, we have found that the permissible ranges for the gas mass and shock velocity are constrained to the reasonable ranges $10^{8}~ M_\odot\sim10^{11}~M_\odot$ and $500~{\rm km\ s^{-1}}\sim40~\rm km\ s^{-1}$, respectively. {\yuan Moreover, a steeper CR distribution with the spectral index $2.1\lesssim s\lesssim2.2$ could be helpful to resolve the tensions between radio and X-ray observations. On the other hand, for NGC 3256, contributions from inverse Compton scattering could be significant since the the core region is compact in the sense of photons. With the constraint $17\ \mu{\rm G}\lesssim B\lesssim33\ \mu\rm G$, our model with a hard spectral index $s\sim2$ can explain the radio and X-ray data simultaneously}. From these two examples, we show that our simple one-zone model can reproduce the radio and X-ray observations of galaxy merger systems. 
Considering the complexity and the diversity observed from system to system, each merging galaxy should be diagnosed independently. We note that since the factor $\frac{1}{2}M_{\rm g}v_s^2$ dominates the electron injections, as can be seen in Equation \ref{eq:elec_injection}, $M_{\rm g}$ and $v_s$ are degenerate in our model. Despite this, our model provides one useful approach to reproduce the radio and X-ray observations and to study the dynamics of galaxy mergers as well as the physical parameters of the shock regions.  

{\yuancc Unavoidably, $pp$ collisions in our model can produce gamma rays through $\pi^0$ decays. In the framework of hadronic process, we estimate the gamma-ray flux from $\pi^0$ decays
\begin{linenomath*}
\begin{equation}\begin{split}
\varepsilon_\gamma F_{\varepsilon_\gamma}(\varepsilon_\gamma)&=\frac{2}{3}\varepsilon_\nu F_{\varepsilon_\nu}(\varepsilon_\nu)|_{\varepsilon_{\gamma}=2\varepsilon_\nu}\\&\lesssim\left(\frac{1}{24\pi d_L^2t_{\rm dyn}}\right)\epsilon_p\mathcal C^{-1}M_{\rm g}v_{\rm s}^2.
\end{split}
\end{equation}
\end{linenomath*}
As for NGC 660, we have $\varepsilon_\gamma F_{\varepsilon_\gamma}\lesssim1.7\times10^{-13}\rm\ erg\ s^{-1}\ cm^{-2}$ while the gamma-ray flux of NGC 3256 satisfies $\varepsilon_\gamma F_{\varepsilon_\gamma}\lesssim2.9\times10^{-13}\rm\ erg\ s^{-1}\ cm^{-2}$. Both of these fluxes are lower than the flux sensitivities of current gamma-ray detectors, such as $Fermi$ LAT\footnote{The Pass 8 sensitivity: \url{https://www.slac.stanford.edu/exp/glast/groups/canda/lat_Performance.htm}}, H.E.S.S \citep{holler2015observations}, MAGIC \citep{aleksic2016major}, HAWC \citep{abeysekara2017observation} and VERITAS \citep{park2015performance}. In the future, the 50-hour sensitivity of the proposed Cherenkov Telescope Array (CTA) in the TeV range can reach $\sim10^{-13}\rm\ erg\ s^{-1}\ cm^{-2}$  \citep{bernlohr2013monte}\footnote{The sensitivity can be also found in \url{http://www.cta-observatory.org/science/cta-performance/}} and our model for the merging galaxies can be further constrained by gamma-ray observations.}

Secondary particle interactions can produce observable emissions not only in interacting galaxy systems but also in star-forming and/or starburst galaxies, where supernovae can accelerate high-energy CRs and trigger subsequent particle interactions. Previous studies incorporating $\pi^0$ decays, bremsstrahlung, inverse Compton and synchrotron emissions have shown that CR interactions can be used to explain the gamma-ray observations of the starburst galaxy M82 \citep{yoast2013winds}, the Cygnus X region \citep{yoast2017gamma1} and the ultra-luminous infrared galaxy Arp 220 \citep{yoast2017gamma2}. Interestingly, for Arp 220 we can estimate the CR luminosity density from a galaxy merger scenario in the central molecular zone as $L_{\rm cr,merger}\simeq\frac{1}{2}\epsilon_p M_{\rm g} v_s^2\left(\frac{R}{v_s}\right)^{-1}\approx9.87\times10^{43}\left(\frac{v_s}{500\rm km\ s^{-1}}\right)^3~\rm erg\ s^{-1}$, using the gas mass $M_{\rm g}=6\times10^8~\rm M_\odot$ \citep{sakamoto2008submillimeter} and $R=70~\rm pc$ \citep{downes2007black}, which is roughly twice as much as the best-fitting supernova CR luminosity \cite{yoast2015cosmic}, $L_{\rm cr,SNe}\simeq E_{\rm cr,SN}{\mathcal R}_{\rm SN}\approx4.76\times10^{43}~\rm erg\ s^{-1}$, for a typical CR energy injected by supernovae of $E_{\rm cr,SN}\approx10^{50}~\rm erg$ and a supernova rate ${\mathcal R}_{\rm SN}\approx 15~\rm yr^{-1}$. This demonstrates that our galaxy merger scenario can fill the gap between the observed gamma-ray flux of Arp 220 and the 2015 gamma-ray prediction from the supernova model \citep[see][]{yoast2015cosmic,yoast2017gamma2}. Even more conservatively, taking the uncertainty in the supernova CR injection energy $5\times10^{49}~{\rm erg\lesssim}E_{\rm cr,SN}\lesssim10^{51}~\rm erg$ \citep{senno2015extragalactic} into consideration, we estimate a luminosity $0.21\lesssim L_{\rm cr,merger}/L_{\rm cr,SN}\lesssim4.15$, which indicates that our model can explain a significant part of the gamma-ray observation.

Various authors, e.g., \cite{thompson2007starburst} and \cite{lacki2014star}, have investigated the contributions from secondary particles (e.g., pions and electrons/positrons) in star-forming/starburst galaxies to the MeV-GeV gamma-ray background and found that these sources can describe a significant portion of the extragalactic gamma-ray background. In this paper, our work has expanded the scope of the applicability of the secondary particle interaction model to galaxy merging systems by introducing a phenomenological approach where CR productions, electron-positron distributions and electromagnetic emissions can be predicted from the basic parameters of the merging regions. This enables us, furthermore, to constrain the gas mass, shock velocity and magnetic field given that supernova CR luminosities and star-formation rates are revealed.

Since galaxy mergers are also promising sources of high-energy neutrinos, these systems may be detected by astrophysical neutrino detectors, such as the IceCube Neutrino Observatory \citep[e.g.,][for reviews]{doi:10.1146/annurev-nucl-102313-025321,halzen2017high}. 
So far, IceCube has detected the diffuse astrophysical high-energy neutrino background \citep{aartsen2013first,icecube2013evidence,aartsen2014observation,aartsen2015combined}, as well as one possible source, blazar TXS 0506+056 \citep{icecube2018neutrino}. 
The physical origin of the bulk of these neutrinos is still under debate, but the success of multi-messenger obswervations following IceCube-170922A show that neutrino astronomy has become an important and indispensable part of multi-messenger astrophysics \citep{Keivani:2018rnh}. 
Our model for high-energy emissions from galaxy mergers connects the electromagnetic emissions from merging regions to the neutrino emission and CR acceleration. 
With the prospects for detecting or setting the limits on their high-energy neutrino emission by current and/or next-generation neutrino detectors \citep{Murase:2016gly,yuan2018cumulative},
our work will be able to  provide a new perspective on future multi-messenger studies of the evolution of galaxies.

\acknowledgements
We are grateful to Shigeo Kimura and Zhao-Wei Zhang for useful discussions. The authors would like to thank the referee for constructive comments and suggestions. This research was partially supported by NASA NNX13AH50G (C.C.Y., P.M.), and the Alfred P. Sloan Foundation and NSF grant PHY-1620777 (K.M.).
\bibliography{kmurase}
\end{document}